\documentstyle[a4,12pt]{report}
\begin{document}
\sloppy

\hfill LYCEN 9419
\vskip 0.07truecm
\hfill Mai 1994

\thispagestyle{empty}

\bigskip
\bigskip
\bigskip
\begin{center}
{\bf \Huge{Supersym\'etrie et Math\'ematiques}}
\end{center}
\bigskip
\bigskip
\centerline{{ {Version \'etendue d'un expos\'e pr\'esent\'e }}}
\centerline{{ {au S\'eminaire d'Analyse de l'Universit\'e Blaise
Pascal }}}
\centerline{{ {(Clermont-Ferrand 2) en mai 1993 }}}
\bigskip
\bigskip
\bigskip
\centerline{\bf Fran\c cois Gieres}
\bigskip
\bigskip
\bigskip
\centerline{{\it Institut de Physique Nucl\'eaire}}
\centerline{\it Universit\'e Claude Bernard (Lyon 1)}
\centerline{\it 43, boulevard du 11 novembre 1918}
\centerline{\it F - 69622 - Villeurbanne C\'edex}
\bigskip
\bigskip
\bigskip
\bigskip

\begin{abstract}
\bigskip
\bigskip

Nous pr\'esentons une introduction aux concepts
de la supersym\'etrie par l'interm\'ediaire de trois
exemples: (i) M\'ecanique quantique supersym\'etrique,
(ii) Superalg\`ebres de Lie, (iii) Superconnexions
de Quillen.
Les points communs \`a toutes ces notions sont soulign\'es et
des applications
sont indiqu\'ees. En particulier nous esquissons la
d\'emonstration du th\'eor\`eme de Gau{\ss} et Bonnet
d'apr\`es Patodi et la d\'emonstration des
in\'egalit\'es de Morse
d'apr\`es Witten.

\end{abstract}
\bigskip

\bigskip
\bigskip

\newcommand{\TH}{\Theta}
\newcommand{\LA}{\Lambda}
\newcommand{\pa}{\partial}
\newcommand{\de}{\delta}
\newcommand{\zz}{{\bf Z}}
\newcommand{\hh}{{\cal H}}
\newcommand{\g}{{\cal G}}
\newcommand{\dd}{{\rm deg}}

\newtheorem{theo}{Th\'eor\`eme}[section]
\newtheorem{lem}{Lemme}[section]
\newtheorem{prop}{Proposition}[section]
\newtheorem{cor}{Corollaire}[section]
\newtheorem{defin}{D\'efinition}[section]

\tableofcontents

\newpage

\setcounter{page}{1}

\chapter{Introduction}

Beaucoup de syst\`emes physiques sont invariants sous un certain
ensemble de
transformations continues qui engendrent
un groupe de Lie.
Ces sym\'etries sont soit des sym\'etries externes
de nature g\'eom\'etrique
(c'est-\`a-dire
li\'ees \`a la structure g\'eom\'etrique de
l'espace-temps), soit des sym\'etries internes
(c'est-\`a-dire
li\'ees aux degr\'es de libert\'e internes comme la charge
\'electrique, l'isospin, ...). Dans tous les cas, l'\'etude
math\'ematique de ces groupes d'invariance, leur introduction
dans les th\'eories physiques et l'exploration de leurs
cons\'equences repr\'esentent des probl\`emes math\'ematiques
int\'eressants.

En fait, ce n'est pas un hasard si l'introduction
et l'\'etude de nouvelles sym\'etries en physique
ont toujours conduit \`a
un rapprochement avec les math\'ematiques
et \`a des interactions tr\`es fructueuses
entre les sciences physiques et math\'ematiques.
Ceci a \'et\'e le cas pour les transformations
de Lorentz et de Poincar\'e en Relativit\'e Restreinte
(Lorentz, Poincar\'e, Einstein, Minkowski, ...), pour
les transformations de coordonn\'ees g\'en\'erales
en Relativit\'e G\'en\'erale (Einstein, Hilbert, Weyl,
Birkhoff, ...) ou pour la repr\'esentation des sym\'etries
en th\'eorie quantique (Weyl, Stone, von Neumann, Wigner,
Bargmann, Mackey, ...). Ces exemples concernent le d\'ebut de
ce si\`ecle et donc une \'epoque o\`u les sciences \'etaient
encore moins ``compartiment\'ees". Vint ensuite la p\'eriode
Bourbaki en math\'ematiques et la poursuite d'une
approche pragmatique en physique (\`a savoir avec le
d\'eveloppement de la th\'eorie des champs qui
a permis de faire des pr\'edictions spectaculairement
pr\'ecises et bien confirm\'ees par l'exp\'erience
tout en reposant sur des fondements math\'ematiques
tr\`es douteux).
Cependant les esprits se sont de nouveaux rapproch\'es durant
la derni\`ere trentaine d'ann\'ees et en partie gr\^ace \`a
l'introduction de nouvelles
sym\'etries. Citons l'exemple des th\'eories de jauge
qui d\'ecrivent les
sym\'etries internes
et pour lesquelles les concepts de base ont \'et\'e d\'evelopp\'e
de mani\`ere
ind\'ependante en math\'ematiques
(Ehresmann, Whitney, Koszul, Chern, ...) et en physique
(Yang, Mills, ...).
Des exemples tr\`es r\'ecents sont ceux des alg\`ebres de sym\'etrie
de dimension infinie et de leurs extensions centrales
(Witt, Virasoro, Gelfand, Kac, Moody, ...), des groupes quantiques
(Kulish, Reshetikhin, Sklyanin, Drinfeld, Jimbo, Woronowicz, ...)
ou des espaces de modules
(Riemann,
Teichm\"uller, ...) qui jouent un r\^ole important dans la th\'eorie
des cordes actuellement discut\'ee en physique.

Et puis il y a l'exemple auquel le pr\'esent expos\'e
est consacr\'e: la {\em supersym\'etrie}.
En gros, ``super" signifie
$\zz_2$-gradu\'e. Cette graduation par le groupe
$\zz_2$ intervient de mani\`ere naturelle
en math\'ematiques, par exemple dans l'alg\`ebre de Grassmann
(alg\`ebre des formes diff\'erentielles) et aussi en physique,
si l'on
consid\`ere les champs spinoriels (comme celui de l'\'electron)
comme des variables grassmanniennes. Ainsi il n'est pas \'etonnnant
que les extensions $\zz_2$-gradu\'ees de l'alg\`ebre lin\'eaire
et de la g\'eom\'etrie diff\'erentielle aient \'et\'e
d\'evelopp\'ees de mani\`ere simultan\'ee et
plus ou moins ind\'ependante par les physiciens et math\'ematiciens.

De la multitude des notions et applications de la supersym\'etrie
nous en avons choisi trois qui sont conceptuellement simples
et qui concernent des domaines diff\'erents.
Elles sont discut\'ees dans les trois chapitres suivants
qui sont essentiellement ind\'ependants entre eux.
En guise de conclusion
nous donnerons un bref aper\c cu
des extensions $\zz_2$-gradu\'ees dans
d'autres domaines des math\'ematiques tout
en mentionnant quelques ouvrages concernant les applications
en physique.

\chapter{M\'ecanique Quantique Supersym\'etrique}

\section{Quelques motivations}

Pour comprendre le formalisme math\'ematique de
la m\'ecanique quantique supersym\'etrique (MQSUSY),
il n'est pas n\'ecessaire de conna\^{\i}tre le
cadre physique correspondant.
N\'eanmoins, pour motiver un peu les d\'efinitions que
nous allons introduire et
les questions que nous allons \'etudier,
nous rappelons d'abord quelques id\'ees physiques concernant
la quantification d'un syst\`eme avec un nombre
fini de degr\'es de libert\'e \cite{mll,jmj}.
A cet effet nous partirons d'un syst\`eme dynamique tr\`es
simple de la m\'ecanique classique \`a une dimension,
\`a savoir l'oscillateur harmonique.

Consid\'erons un point de masse $m\neq 0$ fix\'e \`a un ressort
suppos\'e sans masse:

\vspace{7cm}

Supposons que la position de repos de la masse corresponde
\`a la coordonn\'ee
$x=0$ sur une \'echelle rectiligne. Si on \'etire le ressort, alors
il r\'eagit avec une force de rappel qui est proportionnelle \`a
l'\'elongation, $F=-kx$ ($k$ \'etant une constante
positive). S'il n'y a pas
de friction, l'\'equation de mouvement newtonienne
du point de masse
est donn\'ee par
\begin{equation}
\label{mvt}
m \ddot{x} = -kx
\qquad {\rm avec} \quad
\ddot{x}(t) = \frac{d^2x}{dt^2} (t)
\ \ .
\end{equation}
Elle peut encore
s'\'ecrire comme
\[
\ddot{x} + \omega^2 x =0
\qquad {\rm avec} \quad \omega = \sqrt{\frac{k}{m}}
\ \ .
\]
En multipliant l'\'equation (\ref{mvt}) par $\dot{x}$, nous obtenons
\[
\frac{d\ }{dt} \left( {1\over 2}  m \dot{x} ^2 + {1\over 2}  k x^2
\right) = 0
\ \ .
\]
Donc {\em l'\'energie} m\'ecanique du syst\`eme,
\[
E_{\rm cl} =
{1\over 2} \, m \dot{x} ^2 + V(x)
\qquad {\rm avec} \quad  V(x) = {1\over 2} \, kx^2
\ \ ,
\]
(qui est la somme de l'\'energie cin\'etique et de l'\'energie
potentielle $V(x)$) est conserv\'ee au cours du temps.
Il est utile d'exprimer
cette quantit\'e en fonction des variables
de {\em position} $x$ et {\em d'impulsion}
$p = m\dot{x}$ :
\begin{equation}
\label{ecl}
E_{\rm cl} =
{1\over 2m} \, p^2 +V(x)
\ \ .
\end{equation}
Ainsi, pour une \'energie totale $E_{{\rm cl}} \geq 0$
donn\'ee, on a un mouvement
oscillatoire pour lequel la somme des \'energies cin\'etique et
potentielle est toujours \'egale \`a $E_{\rm cl}$:

\vspace{7cm}

Du point de vue math\'ematique la {\em quantification}
de ce syst\`eme classique revient \`a rempla\c cer
les variables $x, p$ par des op\'erateurs lin\'eaires
$\hat{x}, \hat{p}$ agissant sur un espace de Hilbert
s\'eparable ${\cal H}$. Dans la repr\'esentation dite
de Schr\"odinger, on choisit ${\cal H} = {\cal L}^2 ({\bf R}, dx)$,
c'est-\`a-dire
l'espace de Hilbert associ\'e \`a l'espace vectoriel
complexe
des fonctions $f:{\bf R} \to {\bf C}$ pour lesquelles
$\mid f \mid^2$ est sommable sur ${\bf R}$
par rapport \`a la mesure de Lebesgue.
La variable classique $x$
devient alors l'op\'erateur $\hat{x}$ de multiplication
par $x$ et $p$ devient l'op\'erateur diff\'erentiel
$\hat{p} = -i\hbar \, \pa  /  \pa x$ o\`u $\hbar$ est la constante
de Planck divis\'ee par $2\pi$.
Comme ces op\'erateurs ne sont pas born\'es,
ils ne peuvent pas \^etre d\'efinis sur tout l'espace ${\cal H}$, mais
seulement sur
des sous-espaces denses de ${\cal H}$.
Cependant nous allons ignorer ce point
dans notre illustration.

Alors que $x$ et $p$ commutent entre eux, les op\'erateurs associ\'es
satisfont les {\em relations de commutation canoniques de Heisenberg},
\begin{equation}
\label{ccr}
[ \, \hat{p} \, , \, \hat{x} \, ] \equiv
 \hat{p} \hat{x} -
 \hat{x} \hat{p} = {\hbar \over i} \, {\bf 1}
\ \ ,
\end{equation}
o\`u
${\bf 1}$ d\'enote l'op\'erateur identit\'e sur ${\cal H}$.
Supposons maintenant que les ``fonctions d'onde" $\psi \in
{\cal L}^2 ({\bf R}, dx)$ d\'ependent non seulement de $x$,
mais aussi (de fa\c con param\'etrique et diff\'erentiable)
de la variable de temps $t\in {\bf R}$ :
\begin{eqnarray*}
\psi & : & {\bf R} \times {\bf R} \ \to \ {\bf C}
\\
& & \ (x, t) \ \ \, \mapsto \ \psi (x,t) \equiv \psi_t (x)
\qquad {\rm avec} \quad \psi_t \in {\cal L} ^2 ({\bf R} , dx )
\ \ .
\end{eqnarray*}
Avec l'hypoth\`ese suivant laquelle la variable d'\'energie $E_{\rm cl}$
devient l'op\'erateur
$i\hbar \, \pa /\pa t$
de translation dans le temps dans la th\'eorie
quantique, nous
trouvons alors que l'expression (\ref{ecl}) pour l'\'energie classique
conduit \`a {\em l'\'equation de Schr\"odinger} :
\begin{equation}
\label{sch}
\hat{H} \psi_t = i \hbar \, \frac{\pa \psi_t }{\pa t}
\end{equation}
avec
\begin{equation}
\label{osc}
\hat{H} = {1 \over 2m} \, \hat{p}^2 + V(\hat{x}) = \frac{-\hbar^2}{2m}
\; \frac{\pa^2 \ }{\pa x^2} + {k\over 2} \, \hat{x}^2
\ \ .
\end{equation}
Ceci est l'\'equation d'\'evolution dans le temps
de la th\'eorie quantique et $\hat{H}$
s'appelle {\em l'op\'erateur hamiltonien} ou
le {\em hamiltonien} du syst\`eme
consid\'er\'e.

Int\'eressons nous maintenant
aux {\em \'etats stationnaires}
$\varphi (x)$, c'est \`a dire
\`a des fonctions d'onde correspondant \`a
une \'energie $E\in {\bf R}$ fix\'ee. On les obtient en
substituant
\[
\psi_t (x) = {\rm exp} \left(- \frac{i}{\hbar} Et \right)  \varphi (x)
\]
dans l'\'equation de Schr\"odinger; il s'ensuit que
\begin{equation}
\label{vp}
\hat{H} \varphi = E \varphi
\ \ .
\end{equation}
Ainsi nous sommes amen\'es au probl\`eme math\'ematique suivant:
consid\'erons un op\'erateur lin\'eaire auto-adjoint $\hat{H}$
d\'efini sur un sous-espace dense ${\cal D}(\hat{H})$ d'un espace
de Hilbert s\'eparable ${\cal H}$ et d\'eterminons ses valeurs
propres $E \in {\bf R}$ ainsi que ses fonctions propres
$\varphi \in {\cal D}(\hat{H})$. Pour notre exemple de
l'oscillateur harmonique, la r\'eponse est bien connue:
il existe un nombre d\'enombrable de valeurs propres
et elles sont donn\'ees par
\begin{equation}
\label{el}
E_n = \hbar \omega (n+ {1\over 2} )
\qquad \quad  {\rm avec} \quad
n \in {\bf N}
\ \ .
\end{equation}
Par cons\'equent, un oscillateur harmonique \`a
l'\'echelle d'\'energie $\hbar \omega$ d\'efinie par la constante
de Planck poss\`ede des niveaux d'\'energie quantifi\'es:
seulement des valeurs discr\`etes sont permises pour $E$.
En particulier la plus petite valeur possible est
$E_0= {1\over 2} \hbar \omega$ qui est donc plus grande que z\'ero
(z\'ero correspondrait \`a la position de repos). La vie
dans le monde quantique n'est pas facile, il n'y a point
de repos!

Tous ces d\'etails n'ont d'autres objectifs que de
rappeler des souvenirs et
d'illustrer les cons\'equences physiques dramatiques
de quelques \'equations math\'ematiques plut\^ot
banales et innocentes!

\section{M\'ecanique Quantique Supersym\'etrique}

Comme nous ne reviendrons plus \`a la m\'ecanique
classique dans la suite, nous supprimerons d\'esormais
les chapeaux sur les op\'erateurs.

La MQSUSY consiste dans l'\'etude de syst\`emes physiques
d\'ecrits par des op\'erateurs
hamiltoniens de la forme $H=Q^2$ sur un espace de
Hilbert ${\cal H}$ admettant une d\'ecomposition par $\zz_2$,
c'est-\`a-dire
${\cal H}$ est de la forme $\hh = \hh_b \oplus \hh_f$.
Ces syst\`emes admettent un grand nombre d'applications en
physique et en math\'ematiques. Ici nous allons uniquement
exposer quelques aspects math\'ematiques tout en nous basant
sur la r\'ef\'erence \cite{bs} et tout en renvoyant
le lecteur int\'eress\'e aux travaux \cite{mqs} pour les
applications en physique. Concernant ces derni\`eres nous
remarquons seulement que $H=Q^2$ implique
$[H,Q]=0$, c'est-\`a-dire
$H$ est invariant sous les
transformations g\'en\'er\'ees par $Q$: cette sym\'etrie
de $H$ permet d'expliquer certaines d\'eg\'en\'erescences
dans le spectre de $H$ et elle permet d'appliquer
des m\'ethodes alg\'ebriques pour d\'eterminer ce spectre.

\begin{defin}
Consid\'erons un espace de Hilbert s\'eparable $\hh$.
Soient $H, P, Q$ des op\'erateurs lin\'eaires
auto-adjoints sur $\hh$ et soit $P$ born\'e \footnote{En g\'en\'eral
$H$ et $Q$ ne sont pas born\'es et il n'est donc pas possible
de les d\'efinir sur tout l'espace $\hh$, mais seulement sur des
sous-espaces denses de ${\cal H}$
sur lesquels ces op\'erateurs sont essentiellement
auto-adjoints \cite{rs}.}. On dit que le syst\`eme $(H,P,Q)$
est {\em supersym\'etrique}, si
\begin{eqnarray}
H & = &  Q^2
\nonumber  \\
P^2 & = & {\bf 1}
\label{mq}
\\
\{ Q, P\} & \equiv & QP+PQ =0
\ \ .
\nonumber
\end{eqnarray}
$H$ s'appelle le {\em hamiltonien}, $Q$ {\em l'op\'erateur
de supersym\'etrie} et $P$ {\em l'involution}.
\end{defin}

Les crochets $[ \ , \ ]$ et
$\{ \ , \ \}$ d\'efinis
sur l'alg\`ebre des op\'erateurs par les expressions
(\ref{ccr}) et (\ref{mq}) sont appel\'es
{\em commutateur} et {\em anticommutateur}, respectivement.

Comme $H=Q^2$, un syst\`eme supersym\'etrique est d\'ej\`a
sp\'ecifi\'e par la donn\'ee de $P$ et $Q$; cependant
l'objet d'int\'er\^et principal est le hamiltonien $H$
et il est donc naturel de l'inclure dans la d\'efinition.
La relation de base $H=Q^2$ peut aussi s'\'ecrire comme
\begin{equation}
\label{base}
\{ Q , Q \} =2H
\end{equation}
et c'est pr\'ecisement cette \'equation qui a \'et\'e \`a
l'origine de toutes les th\'eories supersym\'etriques \cite{gl,wz}
et en particulier de la MQSUSY \cite{w1}.
En effet cette relation correspond \`a une repr\'esentation
de la superalg\`ebre de Poincar\'e en dimension d'espace-temps
$0+1$, voir section 3.4.

Les relations (\ref{mq}) ont des cons\'equences multiples que
nous allons \'elaborer maintenant.
Le produit scalaire entre $\varphi, \psi \in {\cal H}$
sera not\'e par $\langle \varphi , \psi \rangle$ et
la norme induite du vecteur $\varphi$ par $\| \varphi \|$.

{\bf (1)} Comme $Q$ est auto-adjoint et $H =Q^2$, on a
\begin{equation}
H \geq 0
\ \ ,
\end{equation}
car $ \langle \varphi , H\varphi \rangle =
\langle Q\varphi , Q\varphi \rangle = \| Q\varphi \|^2 \geq 0$.

{\bf (2)} Comme mentionn\'e en haut, $H =Q^2$ implique
\begin{equation}
[ H , Q ] = 0
\ \ .
\end{equation}
Par ailleurs, de $\{ Q,P\} = 0$ nous d\'eduisons que
\begin{equation}
[ H , P ] = 0
\ \ .
\end{equation}

{\bf (3)} De $P^2 ={\bf 1}$ il suit que l'involution $P$
admet pour seules valeurs propres $\pm 1$. Elle induit
une d\'ecomposition (graduation par $\zz_2$) de l'espace
de Hilbert $\hh$: si $\varphi \in \hh$, alors
\begin{eqnarray*}
\varphi & = &  {1\over 2} \, ( \varphi +P\varphi ) +
{1\over 2} \, ( \varphi -P\varphi )
\\
& \equiv & \varphi_b + \varphi_f
\ \ .
\end{eqnarray*}
Donc,
\begin{equation}
\hh = \hh_b \oplus  \hh_f
\ \ ,
\end{equation}
avec
\begin{eqnarray*}
\hh_b & = & \{ \varphi \in \hh \, \mid \, P\varphi =+ \varphi \}
\\
\hh_f & = & \{ \varphi \in \hh \, \mid \, P\varphi =- \varphi \}
\ \ .
\end{eqnarray*}
Motiv\'e par le r\^ole de l'op\'erateur $Q$ en physique
des particules, on appelle les vecteurs de $\hh_b$ (resp.
$\hh_f$) les {\em \'etats bosoniques} ou {\em pairs}
(resp. {\em fermioniques} ou {\em impairs}).
Avec cette d\'ecomposition l'op\'erateur $P$ s'\'ecrit
suivant
\[
P = \left[
\begin{array}{cc}
{\bf 1}_b & 0 \\
0 & - {\bf 1}_f
\end{array}
\right]
\ \ .
\]

{\bf (4)} L'involution $P$
induit aussi une d\'ecomposition sur l'alg\`ebre des op\'erateurs
agissant sur $\hh$. Soit
\[
K = \left[
\begin{array}{cc}
A & B \\
C & D
\end{array}
\right]
\]
un op\'erateur agissant sur $\hh =\hh_b \oplus \hh_f$.
Alors
\begin{equation}
\label{pa}
[P,K] =0 \qquad \Longleftrightarrow \qquad
K = \left[
\begin{array}{cc}
A & 0 \\
0 & D
\end{array}
\right]
\end{equation}
et
\begin{equation}
\label{odd}
\ \ \{ P,K \} =0 \qquad \Longleftrightarrow \qquad
K = \left[
\begin{array}{cc}
0 & B \\
C & 0
\end{array}
\right]
\ \ .
\end{equation}
En analogie avec la terminologie introduite pour les
\'etats, les op\'erateurs qui commutent avec l'involution
$P$ sont appel\'es {\em op\'erateurs bosoniques} ou {\em pairs}
alors que ceux qui anticommutent avec $P$ sont dits
{\em fermioniques} ou {\em impairs}. Par exemple,
le hamiltonien $H$ est pair et l'op\'erateur de supersym\'etrie
$Q$ impair.

L'involution $P$ sur ${\cal H}$ induit donc une graduation sur
l'alg\`ebre des op\'erateurs lin\'eaires d\'efinis sur ${\cal H}$.
Sur cette alg\`ebre gradu\'ee on peut alors d\'efinir
un commutateur gradu\'e dont le commutateur et l'anticommutateur
consid\'er\'es ci-dessus
sont des cas particuliers, voir section 3.2, \'equation (\ref{scl}).

{\bf (5)} Comme $Q$ est auto-adjoint ($Q^{\ast} =Q$) et
anticommute avec $P$, il suit du raisonnement pr\'ec\'edent
que
\begin{equation}
\label{su}
Q = \left[
\begin{array}{cc}
0 & A^{\ast} \\
A & 0
\end{array}
\right]
\ \ ,
\end{equation}
o\`u $A$ est un op\'erateur lin\'eaire. Appliquons maintenant
$Q$ \`a un vecteur de $\hh$:
\[
Q\varphi
 = \left[
\begin{array}{cc}
0 & A^{\ast} \\
A & 0
\end{array}
\right]
\left[
\begin{array}{c}
\varphi_b \\ \varphi_f
\end{array}
\right]
=
\left[
\begin{array}{c}
A^{\ast} \varphi_f \\ A \varphi_b
\end{array}
\right]
\ \ .
\]
Ceci \'etant de nouveau un vecteur de $\hh_b \oplus \hh_f$,
nous concluons que
\begin{eqnarray}
Q & : & \hh_b \ \to \ \hh_f
\\
Q & : & \hh_f \ \to \ \hh_b
\ \ ,
\nonumber
\end{eqnarray}
ce qui veut dire que
\[
\fbox{\mbox{$Q$ \'echange les \'etats
bosoniques et fermioniques}}
\ .
\]
C'est cette propri\'et\'e
fondamentale de $Q$ qui a motiv\'e la terminologie
op\'erateur de `supersym\'etrie'.

Notons aussi que (\ref{su}) implique que $H$ a la forme
suivante:
\begin{equation}
H = \left[
\begin{array}{cc}
A^{\ast} A & 0 \\
0& A A^{\ast} \\
\end{array}
\right]
\ \ .
\end{equation}

{\bf (6)} Pour conclure, nous en venons \`a la propri\'et\'e
fondamentale de tout syst\`eme supersym\'etrique.
Supposons que
\[
H\varphi = E \varphi
\qquad {\rm avec} \qquad E>0
\ \ .
\]
En appliquant l'op\'erateur $Q$ \`a cette relation et
en utilisant que $[H,Q]=0$, nous trouvons que
\[
H(Q\varphi)  = E (Q\varphi )
\ \ .
\]
Donc, si $\varphi$ est un vecteur propre de $H$, alors
$Q\varphi$ est aussi un vecteur propre pour la m\^eme valeur
propre $E>0$. (Remarquons que ce raisonnement n'est pas
valable pour la valeur propre nulle : la relation $H\varphi =0$
implique
\[
0= \langle \varphi, H\varphi \rangle =
\langle \varphi, Q^2 \varphi \rangle =
\langle Q \varphi, Q\varphi \rangle = \| Q\varphi \|^2
\ \ ,
\]
donc $Q\varphi =0$ et z\'ero n'est pas un vecteur propre
par d\'efinition.)

Comme on l'a montr\'e plus haut, $\varphi \in \hh_b$ (resp. $\hh_f$)
implique
$Q\varphi \in \hh_f$ (resp. $\hh_b$). Ainsi nous
avons d\'eriv\'e le r\'esultat suivant :
\begin{theo}[Propri\'et\'e fondamentale d'un syst. supersym\'etrique]
Pour un syst\`eme supersym\'etrique les valeurs propres {\em non}
nulles du hamiltonien $H$ admettent le m\^eme nombre de
vecteurs propres bosoniques et fermioniques:
\begin{equation}
\label{pf}
{\rm dim\  Ker\ } \left[ (H-E) \lceil \hh_b \right]  =
{\rm dim\  Ker\ } \left[ (H-E) \lceil \hh_f \right]
\ \ .
\end{equation}
\end{theo}
Ici nous avons utilis\'e la notation
\[
\varphi \in {\rm Ker} \, \left[ (H-E) \lceil \hh_i \right]
\quad \Longleftrightarrow \quad
(H-E) \varphi = 0 \quad {\rm et} \quad \varphi \in \hh_i
\quad  (i\in \{ b,f\} )
\ \ .
\]
D'une mani\`ere g\'en\'erale la restriction d'un op\'erateur
lin\'eaire $A$ sur ${\cal H}$ \`a un sous-espace
${\cal G}$ de ${\cal H}$ sera not\'ee par $A \lceil {\cal G}$.

Remarquons que la formule (\ref{pf}) pour le spectre de $H$
peut \^etre pr\'esent\'ee
d'une mani\`ere plus rigoureuse en utilisant les projecteurs
$P_{\Delta}$ sur ${\cal H}$ qui d\'efinissent
la d\'ecomposition spectrale de $H$ (c'est-\`a-dire
$H = \int_{{\bf R}} d\lambda \ \lambda \  P_{\lambda}$) : ainsi
(\ref{pf}) s'\'ecrit
\[
{\rm dim} \ \left[ P_{\Delta} \lceil \hh_b \right]  =
{\rm dim} \ \left[ P_{\Delta} \lceil \hh_f \right]
\]
pour tout sous-ensemble $\Delta$ ouvert et born\'e de l'intervalle
$(0,\infty )$.

\subsection{Exemple : Op\'erateur
de Lapace et Beltrami}

Soit $M$ une vari\'et\'e qui a toutes les bonnes
propri\'et\'es que l'on puisse souhaiter: c'est une vari\'et\'e
r\'eelle de type $C^{\infty}$, de dimension finie $n$, sans bord,
compacte, riemannienne et orient\'ee.

Sur cette vari\'et\'e nous consid\'erons le fibr\'e vectoriel
des formes diff\'erentielles. Soit $\LA^pM$ l'espace vectoriel
des sections de type $C^{\infty}$ dans ce fibr\'e: en terme de
coordonn\'ees locales $(x_1,...,x_n)$ d\'efinies dans un voisinage
d'un point $x \in M$, un \'el\'ement
de $\LA^pM$ est donn\'e par
\[
\alpha (x) = \sum_{1\leq i_1 <...<i_p \leq n}
\alpha _{i_1...i_p} (x) \  dx^{i_1} \wedge ... \wedge dx^{i_p}
\ \ .
\]
Une m\'etrique riemannienne $g$ sur $M$ induit une m\'etrique
$\tilde{g}$ sur $\LA^pM$. En utilisant la norme associ\'ee
\`a $\tilde{g}$ on peut
compl\'eter $\LA^pM$ pour obtenir un espace de Hilbert que nous
d\'enotons
$\overline{\LA^pM}$. Dans la suite nous n'allons pas toujours
explicitement \'ecrire la barre, parce qu'elle n'est pas essentielle
dans la plupart des raisonnements que nous allons faire.

Les op\'erateurs de diff\'erentiation et de
codiff\'erentiation des formes diff\'erentielles sont not\'es par
$d$ et $d^{\ast}$,
\begin{eqnarray*}
d & : &
\LA^p M \ \longrightarrow \ \LA^{p+1}M
\\
d^{\ast} & : & \LA^p M \ \longrightarrow \ \LA^{p-1}M
\ \ ,
\end{eqnarray*}
$d^{\ast}$ \'etant l'adjoint de $d$ par rapport \`a la
m\'etrique $\tilde g$.

Sur $\hh = \oplus_{p=0}^{n} \,
\overline{\LA^pM}$ nous introduisons le {\em syst\`eme
supersym\'etrique de Laplace et Beltrami:}
\begin{eqnarray}
Q & = & d+d^{\ast}
\nonumber  \\
P\lceil \, \overline{\LA^pM} & = & (-1)^p \, {\bf 1}
\label{lb} \\
L & = &  Q^2 = dd^{\ast} +d^{\ast} d
\ \ .
\nonumber
\end{eqnarray}
Le hamiltonien de ce syst\`eme est donc l'op\'erateur
de Laplace et Beltrami $L$ (associ\'e \`a la m\'etrique $g$).
Pour en avoir une id\'ee un peu plus concr\`ete nous rappelons
que dans un syst\`eme de coordonn\'ees locales $(x_1,...,x_n)$
l'action de $L$ sur une fonction $f\in \LA^0M = C^{\infty}(M)$
est donn\'ee par
\[
Lf = \sum_{i,j=1}^n \frac{1}{\sqrt{{\rm det} \, g}} \
\frac{\pa \ }{\pa x_i} \left(
\sqrt{{\rm det} \, g} \, g^{ij} \,
\frac{\pa f }{\pa x_j} \right)
\ \ ,
\]
o\`u $\left( g^{ij} \right)$ et ${\rm det} \, g$ sont respectivement
la matrice
inverse et le d\'eterminant de la matrice avec les \'el\'ements
$g_{ij} = g(
\pa /\pa x_i \, , \,
\pa / \pa x_j )$.

V\'erifions que le syst\`eme (\ref{lb}) satisfait bien toutes
les conditions requises par la d\'efinition 2.1. Il est clair que
$Q,\, P$ et $L$ sont des op\'erateurs auto-adjoints.
Par ailleurs, $P$ est
born\'e
et son carr\'e est l'op\'erateur unit\'e. Pour v\'erifier que
$Q$ anticommute avec $P$, nous choississons $\alpha \in \LA^pM$ et
appliquons $Q$, resp. $P$ sur $\alpha$:
\begin{eqnarray*}
Q\alpha & = & d\alpha + d^{\ast} \alpha \; \in \LA^{p+1}M \oplus
\LA^{p-1}M
\\
P\alpha & = & (-1)^p \, \alpha
\ \ .
\end{eqnarray*}
Ainsi $PQ\alpha = -(-1)^p \, Q\alpha$ et $QP\alpha = (-1)^p \, Q\alpha$,
d'o\`u $(PQ+QP)\alpha =0$.

Dans l'exemple pr\'esent, la d\'ecomposition de l'espace de Hilbert
$\hh$ induite par l'involution $P$ prend la forme
\[
\hh = \hh_b \oplus \hh_f
\qquad \quad {\rm avec} \qquad
\hh_b = \bigoplus_{p\; {\rm pair}} \; \overline{\LA^pM}
\quad , \quad
\hh_f = \bigoplus_{p\; {\rm impair}} \; \overline{\LA^pM}
\ .
\]
L'application du th\'eor\`eme 2.2.1 \`a ce syst\`eme supersym\'etrique
donne le r\'esultat suivant. Pour toute valeur propre $E\geq 0$
de $L$, notons
\begin{eqnarray}
\label{mult}
M_p (E) &=& {\rm dim \; Ker} \left[ (L-E) \lceil (\LA^pM) \right]
\\
& = & \mbox{ multiplicit\'e  de  la  valeur  propre $E$ de
$L$ sur $\LA^pM$}
\ \ .
\nonumber
\end{eqnarray}
Alors l'\'equation (\ref{pf}) implique
\[
\sum_{p \; {\rm pair}} M_p(E) =
\sum_{p \; {\rm impair}} M_p(E)
\qquad \quad {\rm pour} \ E>0
\ \ ,
\]
c'est-\`a-dire
\begin{equation}
\label{pf1}
\sum_{p=0}^n (-1)^p \, M_p(E) = 0
\qquad \quad {\rm pour} \ E>0
\ \ .
\end{equation}
Dans les prochaines sections nous allons voir comment cette
relation peut \^etre directement ou indirectement utilis\'ee
pour d\'emonter les th\'eor\`emes d'indice (Gau{\ss} et Bonnet,
Morse, ...) ou les in\'egalit\'es de Morse.

\section{Applications}
\subsection{Quelques rappels (Betti, de Rham, Hodge, Euler)}

Les {\em nombres de Betti} $b_p \equiv b_p (M)$
$(0 \leq p \leq n )$ d'une vari\'et\'e $M$ de dimension $n$
sont des invariants topologiques d\'efinis comme
les dimensions des groupes d'homologie de $M$ \cite{st}.
D'apr\`es le th\'eor\`eme de de Rham, les groupes
d'homologie et de cohomologie sont isomorphes et ont donc
la m\^eme dimension; ainsi
\[
b_p = {\rm dim} \, H^p (M)
\qquad
(0 \leq p \leq n )
\ \ ,
\]
o\`u $H^p (M)$ sont les groupes de cohomologie de de Rham:
\[
H^p (M) \; = \; \frac{{\rm Ker} \, d}{{\rm Im} \, d} \; = \;
\frac{\{
\alpha \in \LA^p M \; \mid \;   d\alpha =0\} }{ \{
\alpha \in \LA^p M \; \mid \; \alpha = d\beta \ {\rm avec} \ \beta \in
\LA^{p-1} M \} }
\ \ .
\]
La th\'eorie de Hodge implique que la dimension de $H^p (M)$ est
la m\^eme que celle de l'espace des {\em $p$-formes harmoniques} sur
$M$, c'est-\`a-dire
de l'espace des $p$-formes annihil\'ees par l'op\'erateur
de
Laplace et Beltrami $L$:
\begin{equation}
\label{har}
b_p = {\rm dim} \, H^p (M)
= {\rm dim \; Ker} \, \left[ L \lceil \LA^p M \right]
\ \ .
\end{equation}
(Concernant la d\'emonstration de ce r\'esultat nous
remarquons seulement que $L\alpha \equiv (d d^{\ast} +d^{\ast}d)
\alpha =0$ si et seulement si
$d\alpha =0=d^{\ast} \alpha$, car
$\langle \alpha , L\alpha \rangle = \| d\alpha \| ^2 + \| d^{\ast}
\alpha \| ^2$.)

Dans les prochaines sections nous ferons aussi appel
\`a la {\em caract\'eristique d'Euler} $\chi (M)$ de la vari\'et\'e
$M$ d\'efinie par
\begin{equation}
\label{car}
\chi (M) = \sum_{p=0}^{n} (-1)^p b_p
\ \ .
\end{equation}
A titre d'exemple, citons les vari\'et\'es compactes
et orientables de dimension $2$ qui sont toutes diff\'eomorphes
\`a une sph\`ere avec un certain nombre $g$ de trous ($g$ = genre
de $M$) :
\begin{equation}
\label{d2}
b_0 = 1 = b_2 \quad , \quad b_1 = 2g \qquad \Rightarrow \qquad
\chi (M) = 2 - 2g
\ \ .
\end{equation}

\subsection{D\'emonstr. du th\'eor\`eme de Gau{\ss} et Bonnet
d'apr\`es Patodi}

Dans son travail sur les in\'egalit\'es de Morse \cite{w2},
Witten a indiqu\'e des relations entre le th\'eor\`eme
d'indice d'Atiyah-Singer et la MQSUSY. Ces remarques furent
exploit\'ees par d'autres physiciens, notamment
Alvarez-Gaum\'e, Friedan et Windey \cite{ag},
pour donner une d\'emonstration simple du th\'eor\`eme
d'indice; des versions rigoureuses de ces preuves ont \'et\'e
fournies par Getzler et Bismut \cite{ge}.

Dans la suite
nous indiquerons de quelle mani\`ere la
supersym\'etrie intervient dans ces d\'emonstrations
en choississant comme illustration
l'exemple tr\`es simple du th\'eor\`eme de
Gau{\ss} et Bonnet classique \cite{bs}:

\begin{theo}[Gau{\ss} et Bonnet]
Soit $M$ une vari\'et\'e riemannienne compacte, orientable,
de dimension $n=2$. Alors la caract\'eristique d'Euler
et le scalaire de courbure $R$ de $M$ sont reli\'es par
\begin{equation}
\label{gb}
\chi (M) = \frac{1}{4 \pi} \int_M R(x) dx
\end{equation}
ou, d'apr\`es l'\'equation (\ref{d2}),
\[
\int_M R(x) dx  = 8 \pi (1- g )
\qquad (\,  g = {\rm genre \ de}\  M \, )
\ \ .
\]
\end{theo}

Notons \`a ce sujet que la formule (\ref{gb}) peut \^etre
g\'en\'eralis\'ee \`a des vari\'et\'es de dimension
paire $n=2k$ en rempla\c cant la $2$-forme
$(1/4\pi) R(x)dx$ par la $n$-forme d'Euler $E(x)dx$;
la d\'emonstration indiqu\'ee ci-dessous
(qui est due \`a Patodi \cite{p,bs})
s'applique aussi \`a ce cas plus g\'en\'eral.

Esquisse de {\em d\'emonstration}:
La premi\`ere \'etape de la d\'erivation de (\ref{gb})
consiste \`a utiliser la propri\'et\'e de supersym\'etrie
(\ref{pf1}) pour prouver la {\em formule de McKean et Singer}
\cite{mcs} (formule valable pour toute vari\'et\'e compacte
et orientable $M$):
\begin{equation}
\label{mcs}
\chi (M) = {\rm Str} \,  e^{-t L}
\qquad (t\in {\bf R})
\ \ .
\end{equation}
Ici $L$ est l'op\'erateur de Laplace et Beltrami et
`Str' est la {\em supertrace} d\'efinie par\footnote{Pour
les d\'etails d'analyse nous renvoyons \`a la section
12.3 de \cite{bs}.}
\begin{equation}
\label{tro}
{\rm Str }\, e^{-tL} = \sum_{p=0}^{n} (-1)^p \,
{\rm Tr} \, e^{-tL_p}
\qquad {\rm avec} \quad L_p \equiv L \lceil \LA^p M
\ \ .
\end{equation}

D\'erivons maintenant la formule
(\ref{mcs}).
D'apr\`es la d\'efinition (\ref{mult}) de $M_p(E)$,
nous avons
\begin{eqnarray*}
{\rm Str }\, e^{-tL} &=& \sum_{p=0}^{n} (-1)^p \,
{\rm Tr} \, e^{-tL_p}
\\
&=& \sum_{p=0}^{n} (-1)^p
\left[ \sum_E M_p(E)  e^{-tE} \right]
\ \ .
\end{eqnarray*}
Comme l'op\'erateur
$e^{-tL_p}$ est de trace finie, nous pouvons interchanger
les sommes et ensuite appliquer l'\'equation (\ref{pf1}):
\begin{eqnarray*}
{\rm Str }\, e^{-tL} &=& \sum_E e^{-tE} \left[ \sum_{p=0}^{n} (-1)^p
M_p(E) \right]
\\
&=& \sum_{p=0}^{n} (-1)^p M_p(0)
\ \ .
\end{eqnarray*}
Comme $M_p (0) = {\rm dim \; Ker} \left[ L \lceil \LA^p M
\right] = b_p$ (voir \'equation (\ref{har})),
nous obtenons le r\'esultat (\ref{mcs}).

La seconde \'etape de la d\'emonstration de (\ref{gb})
consiste \`a introduire le noyau int\'egral
$e^{-tL_p} (x,y)$
associ\'e \`a $e^{-tL_p}$,
\[
\left( e^{-tL_p} u \right)(x) = \int_M e^{-tL_p} (x,y)
\, u(y) \, dy
\]
et \`a relier le noyau de ${\rm Str} \, e^{-tL}$
au scalaire de courbure $E(x) = (1/4\pi) R(x)$
pour $n=2$ (ou plus g\'en\'eralement \`a la $n$-forme
d'Euler, $E(x) dx$, pour $n=2k$). Cette partie n\'ecessite une
\'etude
analytique plus approfondie qui a \'et\'e faite par Patodi \cite{p}
et qui est d\'ecrite en d\'etail
dans l'ouvrage \cite{bs}.
La supersym\'etrie intervient dans cette
partie par l'interm\'ediaire de la formule de Berezin et Patodi
pour la supertrace.
\hfill $\Box$

\bigskip

Il est \'evident que la d\'emarche suivie dans cette d\'emonstration est
tr\`es diff\'erente de celle suivie dans la d\'erivation habituelle
de la formule de Gau{\ss} et Bonnet \cite{st}: cette derni\`ere
fait appel \`a une triangulation de la vari\'et\'e et
\`a l'interpr\'etation de la courbure en fonction du
transport parall\`ele le long d'une courbe ferm\'ee.

\subsection{D\'emonstr. des in\'egalit\'es de Morse d'apr\`es Witten}

\subsubsection{Th\'eorie de Morse}

Comme notre but ne consiste pas \`a  pr\'esenter les r\'esultats
les plus forts \`a partir des hypoth\`eses les plus faibles,
mais plut\^ot \`a illustrer les id\'ees et m\'ethodes,
nous consid\'erons dans la suite une vari\'et\'e $M$ de type $C^{\infty}$
qui poss\`ede toutes les propri\'et\'es mentionn\'ees au d\'ebut
du chapitre 2.2.1.

Soit $f: M \to {\bf R}$ une fonction de type $C^{\infty}$ sur $M$.
La topologie de $M$ impose des restrictions sur le comportement
de $f$. Plus sp\'ecifiquement les in\'egalit\'es de Morse
donnent une limite inf\'erieure pour le nombre de points critiques de $f$
en fonction de la
topologie de $M$. Avant de formuler ces restrictions,
nous rappelons d'abord
la d\'efinition des notions dont nous aurons besoin
\cite{bs,m}.

\begin{defin}
Soit
$f:M \to {\bf R}$ une fonction de type $C^{\infty}$ et $(x_1,...,x_n)$
un syst\`eme de coordonn\'ees locales sur $M$.

(i) Un point $m\in M$ est un {\em point critique} de $f$, si
\[
df(m) =0 \qquad \Longleftrightarrow \qquad
\frac{\pa f}{\pa x_1} (m) = ... =
\frac{\pa f}{\pa x_n} (m) = 0
\ \ .
\]

(ii) Un point critique $m$ de $f$ est {\em non-d\'eg\'en\'er\'e},
si la hessienne $A$ de $f$ au point $m$ est non-d\'eg\'en\'er\'ee,
c.\`a.d. ${\rm det} \, A(m) \neq 0$;
localement $A(m)$ est donn\'ee par la matrice r\'eelle et sym\'etrique
\[
A(m) =  \left(
\frac{1}{2} \ \frac{\pa^2 f}{\pa x_i \, \pa x_j} (m)  \right)
\ \ .
\]

(iii) {\em L'indice} d'un point critique $m$ de $f$ est le nombre
de valeurs propres n\'egatives de la matrice $A(m)$.
\end{defin}

Remarquons
qu'un point critique non-d\'eg\'en\'er\'e est n\'ecessairement
isol\'e, c'est-\`a-dire
il existe un voisinage de ce point critique
qui ne contient pas d'autres points critiques \cite{m}. Par ailleurs,
si le point critique est
non-d\'eg\'en\'er\'e et poss\`ede un indice `${\rm ind} \,m$', alors
la fonction $f$ prend la forme suivante en fonction de coordonn\'ees
locales $x=(x_1,...,x_n)$ d\'efinies dans un voisinage de $m$ :
\begin{equation}
\label{mor}
f(x) = f(m) + (x_1)^2 + ...+ (x_{n-{\rm ind}\, m})^2
- (x_{n-{\rm ind}\, m +1})^2 -...-(x_n)^2
\ \ .
\end{equation}
Ces coordonn\'ees s'appellent les {\em coordonn\'ees de Morse}.

\begin{defin}
Une fonction $f:M \to {\bf R}$ de type $C^{\infty}$
est appel\'ee {\em fonction de Morse}, si
elle admet un nombre fini de points critiques et
que tous ceux-ci sont non-d\'eg\'en\'er\'es.
Pour une telle fonction on note
\[
m_p(f) = \mbox{multiplicit\'e (nombre) des
points critiques de $f$ avec indice $p$}
\]
\[
(\, 0 \leq p \leq n= {\rm dim}\, M\, )
\ \ .
\]
\end{defin}

Pour illustrer ces d\'efinitions nous consid\'erons l'exemple
du tore bidimensionnel, $M= T^2 =S^1 \times S^1$ \cite{bs}.
On peut plonger cette vari\'et\'e dans ${\bf R}^3$ en associant
\`a tout point $m \in M$ les coordonn\'ees $(m_1, m_2, m_3)
\in {\bf R}^3$:
\[
M \ni m \quad \leftrightarrow \quad (m_1, m_2, m_3)
\in {\bf R}^3
\ \ .
\]
Comme fonction sur $M$ nous choisissons la fonction `hauteur'
d\'efinie par
\begin{eqnarray*}
f & : & M \ \to \ {\bf R}
\\
& & m \ \mapsto \ m_3
\ \ .
\end{eqnarray*}

\vspace{7cm}

Cette fonction admet 4 points critiques, les indices et multiplicit\'es
\'etant les suivants:
\begin{eqnarray*}
{\rm indice} \ p=0 \ :\qquad  m_0(f) &=& 1 \quad \leftrightarrow \quad
\mbox{minimum local de $f$}
\\
{\rm indice} \ p=1 \ :\qquad  m_1(f) &=& 2 \quad \leftrightarrow \quad
\mbox{point de selle local de $f$}
\\
{\rm indice} \ p=2 \ :\qquad  m_2(f) &=& 1 \quad \leftrightarrow \quad
\mbox{maximum local de $f$}
\ \ .
\end{eqnarray*}
Comme le tore est une surface de genre $g=1$,
ses nombres de
Betti sont donn\'es par $b_0 = 1 = b_2$ et $b_1 =2$. Nous avons donc
les \'egalit\'es suivantes dans notre exemple:
\[
m_p(f) = b_p \qquad (0\leq p \leq 2)
\ \ .
\]
Dans le cas g\'en\'eral on a les in\'egalit\'es suivantes qui
imposent une restriction sur les nombres
$m_p(f)$ en fonction de la topologie de $M$ caract\'eris\'ee par
les coefficients $b_p$:

\begin{theo}[In\'egalit\'es de Morse faibles]
Pour toute fonction de Morse $f$ sur $M$ on a
\begin{equation}
\label{imfa}
m_p(f) \geq b_p
\qquad (0\leq p \leq n)
\ \ .
\end{equation}
\end{theo}

Ces relations sont une cons\'equence du r\'esultat suivant.

\begin{theo}[In\'egalit\'es de Morse fortes]
Pour toute fonction de Morse $f$ sur $M$ on a
les in\'egalit\'es
\begin{equation}
\label{imf}
\sum_{p=0}^{k} (-1)^{k-p} m_p(f) \geq
\sum_{p=0}^{k} (-1)^{k-p} b_p
\qquad (0 \leq k \leq n)
\ \ .
\end{equation}
Pour $k=n$, on a l'\'egalit\'e
\begin{equation}
\sum_{p=0}^{n} (-1)^{p} m_p(f) =
\sum_{p=0}^{n} (-1)^{p} b_p
\quad ( = \chi (M) )
\ \ ,
\end{equation}
qui peut encore s'\'ecrire sous la forme
\begin{equation}
\label{tm}
\sum_{p=0}^{n} (-1)^{p} \left[ b_p - m_p(f) \right] = 0
\end{equation}
et qui s'appelle le {\em th\'eor\`eme d'indice de Morse}.
\end{theo}

En ajoutant les in\'egalit\'es fortes (\ref{imf}) pour $k$
et $k-1$, on obtient les in\'egalit\'es faibles (\ref{imfa}).

\subsubsection{D\'emonstration d'apr\`es Witten}

La d\'emonstration classique du th\'eor\`eme pr\'ec\'edent \cite{m}
part de la d\'efinition topologique des nombres de Betti
(en tant que dimensions des groupes d'homologie de $M$) et
utilise des r\'esultats g\'en\'eraux de la th\'eorie d'homologie
(concernant les suites
exactes des groupes d'homologie relatifs et les liens avec
l'homotopie).
Par contre
la d\'emonstration de Witten \cite{w2,hen} part de la d\'efinition
analytique du nombre de Betti $b_p$ (en tant que dimension de l'espace
des $p$-formes harmoniques sur $M$).
A c\^ot\'e de son caract\`ere original, elle pr\'esente
un int\'er\^et pour diff\'erentes raisons.
D'abord elle \'etablit des connexions int\'eressantes entre
la g\'eom\'etrie et l'analyse. Ensuite, la transposition
des arguments de Witten dans le cadre de la g\'eom\'etrie
alg\'ebrique sur un corps fini a permis \`a Laumon
de d\'emontrer un r\'esultat jusqu'alors conjectural
(concernant la formule du produit pour la constante
de l'\'equation fonctionnelle de la fonction $L$
attach\'ee \`a une repr\'esentation $l$-adique)
\cite{lau,hen}. Finalement
le syst\`eme supersym\'etrique introduit dans la
d\'emonstration de Witten a \'et\'e repris dans la suite par Witten
dans le cadre des th\'eories de Floer et de Donaldson traitant des
invariants topologiques associ\'es aux vari\'et\'es
de dimension $3$ et $4$
(voir r\'ef\'erence \cite{blau} pour une revue de ces r\'esultats).

L'id\'ee de d\'epart de Witten pour d\'eriver
les relations (\ref{imf}) et (\ref{tm}) consiste \`a d\'eformer
le syst\`eme supersym\'etrique associ\'e \`a l'op\'erateur
de Laplace et Beltrami par l'interm\'ediaire d'une fonction de Morse.

Soit $M$ comme dans la section pr\'ec\'edente et
$f:M\to {\bf R}$ une fonction de
Morse. Pour $t\in {\bf R}$ on consid\`ere
\[
d_t = e^{-tf} d e^{tf} \quad , \quad d^{\ast} _t
= e^{tf} d^{\ast} e^{-tf}
\ \ .
\]
Le {\em syst\`eme supersym\'etrique de Witten} est alors d\'efini
sur l'espace de Hilbert ${\cal H} = \bigoplus_{p=0}^{n}
\overline{\LA^pM}$ par
\begin{eqnarray}
Q_t & = & d_t + d_t^{\ast}
\nonumber \\
L_t & = & Q_t^2 \ = \  d_t d_t^{\ast} + d_t ^{\ast} d_t
\label{w}
\\
P \lceil \, \overline{\LA^p M} & = & (-1)^p \, {\bf 1}
\ \ .
\nonumber
\end{eqnarray}
Le th\'eor\`eme 2.2.1 appliqu\'e \`a ce syst\`eme prend de nouveau la
forme (\ref{pf1}),
\begin{equation}
\label{susy}
\sum_{p=0}^{n} (-1)^{p} \,  M_p(E)  = 0  \qquad {\rm pour} \quad E>0
\ \ ,
\end{equation}
o\`u $M_p(E)$ repr\'esente maintenant la multiplicit\'e de la valeur
propre $E$ de $L_t$ sur $\LA^pM$:
\begin{equation}
\label{mul}
M_p(E)  = \mbox{dim Ker} \left[ (L_t -E) \lceil \LA^pM \right]
\ \ .
\end{equation}
D'un autre c\^ot\'e la g\'en\'eralisation des arguments de Hodge
au laplacien d\'eform\'e (\ref{w}) implique que les nombres
de Betti sont donn\'es par
\begin{equation}
b_p =
\mbox{dim Ker} \left[ L_t  \lceil \LA^pM \right]
\ \ .
\end{equation}
D'apr\`es la d\'efinition (\ref{mul}) nous avons donc
\[
b_p = M_p(0)
\]
et le th\'eor\`eme de Morse (\ref{tm}) peut s'\'ecrire
comme
\begin{equation}
\label{tm1}
\sum_{p=0}^{n} (-1)^{p} \left[ M_p(0) - m_p(f) \right] =0
\ \ .
\end{equation}
Ainsi la relation qu'on veut d\'emontrer a une forme semblable \`a
la formule (\ref{susy}) r\'esultant de la supersym\'etrie.

D'abord un long calcul montre que
$L_t$ peut s'exprimer en fonction de $L$
et de $f$ comme
\[
L_t = L + t^2 \| df \|^2 + tA
\ \ ,
\]
o\`u $A$ est un op\'erateur d'ordre z\'ero.
Ensuite on introduit un syst\`eme de coordonn\'ees de Morse
(\ref{mor}) pour repr\'esenter la fonction de Morse $f$
dans un voisinage
du point critique $m$ de $f$
avec indice ${\rm ind} \, m$.
Substituant cette expression dans la formule pr\'ec\'edente
pour $L_t$ on obtient
\[
L_t = -\Delta +4t^2 x^2 + tB
\qquad {\rm avec} \quad x^2 = x_1^2 +...+x_n^2
\ \ .
\]
Ici $\Delta$ est le Laplacien ordinaire
agissant sur les $p$-formes selon
\[
\Delta \left( \sum_{i_1,...,i_p} \omega_{i_1...i_p}
dx_{i_1} \wedge ...\wedge dx_{i_p} \right) =
\sum_{i_1,...,i_p}
\left (\Delta \omega_{i_1...i_p}  \right) \
dx_{i_1} \wedge ...\wedge dx_{i_p}
\]
et $B$ est un op\'erateur qui s'\'ecrit en fonction
d'op\'erateurs de cr\'eation et d'annihilation fermioniques
(familiers aux physiciens). A des facteurs pr\`es, l'op\'erateur
scalaire
$-\Delta +4 x^2$ n'est autre que le hamiltonien de l'oscillateur
harmonique (\ref{osc}) \`a $n$ dimensions dont le spectre
est bien connu (voir \'eq.(\ref{el})).

La d\'erivation de la formule
(\ref{tm1}) et des in\'egalit\'es (\ref{imf}) consiste
maintenant dans une
\'etude assez subtile du spectre de $L_t$ au voisinage de $E=0$.
A cette fin, la
dimension de ${\rm Ker\ } [L_t \lceil \Lambda ^p M ]$
est estim\'ee pour des grandes valeurs du param\`etre $t$
suite \`a l'estimation des valeurs propres de $L_t$ pour
$t \to \infty$.
Nous renvoyons \`a \cite{bs} pour les d\'etails concernant cette
analyse.
\hfill $\Box$

\chapter{Superalg\`ebres de Lie}

Dans notre introduction aux concepts de la supersym\'etrie,
nous aurions peut-\^etre d\^u commencer avec les superalg\`ebres
de Lie pour plusieurs raisons:

\noindent
(i) Cette notion est conceptuellement simple et naturelle.

\noindent
(ii) Du point de vue historique
c'\'etait probablement le premier concept autour de la supersym\'etrie
qui soit apparu en math\'ematiques et en physique.

\noindent
(iii) Ces alg\`ebres interviennent dans des domaines tr\`es
vari\'es des math\'ematiques (par exemple
dans les th\'eories d'homotopie,
de cohomologie, de d\'eformation, ...) et de la physique
(physique des particules \'el\'ementaires, physique nucl\'eaire, ...).

La terminologie employ\'ee dans la lit\'erature n'est pas toujours
uniforme et au lieu de superalg\`ebres de Lie, certains auteurs
parlent d'alg\`ebres de Lie gradu\'ees (ces derni\`eres ayant
encore une autre signification pour une troisi\`eme classe d'auteurs).
La r\'ef\'erence standard pour la th\'eorie des superalg\`ebres de
Lie et de leurs repr\'esentations - r\'ef\'erence
que nous allons suivre au d\'ebut de ce chapitre -
est un article d\'etaill\'e de Kac \cite{vk} (voir aussi
\cite{vik}).
D'autres trait\'es d'introduction \`a ce sujet sont
\cite{cns,scheu,bdw,cor}.
Pour des aspects g\'eom\'etriques (comme la th\'eorie
de Borel,Weil et Bott) nous renvoyons \`a
\cite{kos,pen}, pour les superalg\`ebres de
dimension infinie \`a \cite{clr} et aux
r\'ef\'erences qui sont cit\'ees dans ces articles. Ici
nous nous bornons \`a donner la d\'efinition d'une superalg\`ebre
de Lie et d'en pr\'esenter
quelques exemples et r\'ealisations.
D'autres exemples et applications peuvent \^etre trouv\'es
dans les ouvrages cit\'es, en particulier dans
\cite{vk,cns,bdw,jlk,sor}.

\section{D\'efinition}

Les (super)alg\`ebres de Lie peuvent \^etre d\'efinies sur des corps
assez g\'en\'eraux, mais pour notre illustration nous nous bornons
au corps des nombres r\'eels.
Rappelons d'abord qu'une {\em alg\`ebre de Lie} $\g$ sur ${\bf R}$
est un espace vectoriel r\'eel muni d'une op\'eration
(appel\'ee crochet de Lie),
\begin{eqnarray*}
[ \ , \ ] & : & \g \times \g \ \, \to \ \, \g
\\
& & \, (x,y) \  \, \mapsto \ [x,y]
\ \ ,
\end{eqnarray*}
satisfaisant les propri\'et\'es suivantes:

\noindent (i) $[\ , \ ]$ est ${\bf R}$-bilin\'eaire,

\noindent (ii) $[\ , \ ]$ est antisym\'etrique:
$[x,y]= -[y,x]$,

\noindent (iii) $[\ , \ ]$ satisfait l'identit\'e de Jacobi:
$0= [x,[y,z]] +\mbox{permutations circulaires}$.

La g\'en\'eralisation
supersym\'etrique de l'alg\`ebre de Lie consiste \`a
introduire une d\'ecomposition par $\zz_2$ sur l'espace vectoriel $\g$
et \`a incorporer cette graduation dans les
propri\'et\'es (ii) et (iii).
Comme nous n'allons pas discuter une graduation par $\zz$ ici, mais
seulement par $\zz_2$, nous notons les \'el\'ements
(classes d'\'equivalence)
de $\zz_2$ simplement par $0$ et $1$.

\begin{defin} Une {\em superalg\`ebre de Lie}
(ou {\em alg\`ebre de Lie gradu\'ee})
${\cal G}$ sur ${\bf R}$
est donn\'ee par un espace vectoriel r\'eel qui est $\zz_2$-gradu\'e,
\[
{\cal G} = {\cal G}_0 \oplus {\cal G}_1
\ \ ,
\]
et une op\'eration
(appel\'ee supercrochet de Lie), $[\ , \ \} :\g \times \g \to
\g$, satisfaisant les axiomes suivants:

\noindent {\rm (c)} $[\ , \ \}$ est compatible avec la graduation,
c'est-\`a-dire
\begin{equation}
\label{com}
[ \g_k , \g_l \} \subset \g_{k+l}
\qquad \quad {\rm pour} \quad k,l\in \{ 0,1\}
\ \ ,
\end{equation}
ou plus explicitement,
\begin{eqnarray*}
{[ \g _0 , \g _0 \} } & \subset & \g _0
\\
{[ \g _0 , \g _1 \} } & \subset & \g _1
\\
{[ \g _1 , \g _1 \} } & \subset & \g _0
\ \ .
\end{eqnarray*}

\noindent {\rm (i)} $[\ , \ \}$ est ${\bf R}$-bilin\'eaire.

\noindent {\rm (ii)} $[\ , \ \}$ est gradu\'e antisym\'etrique, c.\`a.d.
\begin{equation}
\label{cg}
[x,y \} = -(-1)^{(\dd \, x)(\dd \, y)} [y,x \}
\qquad \quad {\rm avec} \quad
\dd \, x = \left\{
\begin{array}{c}
0 \quad {\rm si} \ \, x\in \g_0 \\
1 \quad {\rm si} \ \, x\in \g_1
\end{array}
\right.
\ \ ,
\end{equation}
et, pour $x_k, y_k \in \g_k$, on a donc
\begin{eqnarray}
{[x_0,y_0 \} } & = & -  {[y_0 ,x_0 \} }
\nonumber
\\
{[x_0 ,y_1 \} } & = & - { [y_1 ,x_0 \} }
\label{as}
\\
{[x_1 ,y_1 \} } & = & \ \ \, { [y_1 ,x_1 \} }
\ \ .
\nonumber
\end{eqnarray}

\noindent {\rm (iii)} $[\ , \ \}$ satisfait l'identit\'e de Jacobi
gradu\'ee,
c'est-\`a-dire
$0= [x,[y,z\} \} $ plus
permutations circulaires avec signes appropri\'es:
\begin{equation}
0= [x,[y,z\} \}
+ (-1)^{(\dd \, z)(\dd \, x + \dd \, y)} [z,[x,y \} \}
+ (-1)^{(\dd \, x)(\dd \, y + \dd \, z)} [y,[z,x \} \}